\documentclass[iop]{emulateapj-rtx4} 
\shortauthors{Sekanina}
\shorttitle{Phase Law and Light Curve of Comet C/2023 A3}
\slugcomment{Version \today }

\begin{document}
\title{A Note on the Phase Law and Light Curve of Comet Tsuchinshan-ATLAS
 (C/2023 A3)\\[-1.6cm]}
\author{Zdenek Sekanina}
\affil{La Canada Flintridge, California 91011, U.S.A.; {\sl ZdenSek@gmail.com}}
\begin{abstract} 
The light curve of comet Tsuchinshan-ATLAS peaked in mid-April 2024,
which nearly coincided with a minimum phase angle of 2$^\circ\!$.9.
The question of a possible correlation between the two events has
implications for the comet's overall performance.  In this note 
I examine the light curve at times of equal phase angles to circumvent
the effect and show that the comet was as bright intrinsically in
late March as it was in early~May.
From a plot of the comet's magnitude~at~unit~geocentric~distance
against the phase angle before and after its minimum on 18~April I derive
a very steep phase~law~and~a~relatively flat, $r^{-2.55}$ light curve
failing to fit the recently reported magnitudes.  After stalling in May
and June, the comet's activity enters another surge (at a time of peak
phase effect), as fragmentation continues.

\end{abstract}

\section{Introduction} 
The approximate coincidence in mid-April 2024 between the timing
of the minimum phase angle of comet Tsuchinshan-ATLAS on the one
hand and the peaks on its light curve and the curve of the dust
production proxy {\it Af}$\rho$ on the other hand is a controversial
issue, because the correlation may or may not be fortuitous.
Even~if~it~is not, one is by no means certain that the phase effect
could account for the peak in its entirety.  In my recent investigation
of the comet (Sekanina 2024, referred to hereafter as Paper~1) the
phase effect was ignored in part because the nature of its source was
not obvious and it was unclear what kind of a phase law
to apply.

Scattering properties of micron- and submicron-sized dust grains
are normally the main factor influencing the brightness of a
cometary atmosphere that needs to be corrected for using one of the
existing methods, such as the model proposed by Marcus (2007).  However,
comet Tsuchinshan-ATLAS fails to display a tail consisting of microscopic
dust, as demonstrated in Paper~1, and one should question the existence
of these tiny grains, in nontrivial amounts, in the coma as well.

Measurements of {\it Af}$\rho$, near 2300~cm in late June and the first
half of July\footnote{See {\tt http://astrosurf.com/cometas-obs.}}, down
from \mbox{6000--9000}~cm in mid-April, show that the production of dust
is high, but that the dust seen in the tail was emitted at times comparable
to the comet's discovery time and has been subjected to solar radiation
pressure accelerations not exceeding 1~percent of the solar gravitational
acceleration.  The relevant particles are submillimeter-sized and larger.

\section{Investigating the Light Curve Near Its Peak}
In the absence of any information on the phase law that affects the
comet's light curve, one can circumvent the problem by employing a
method introduced in the following, if the same phase angle happens to
occur at more than a single point along the orbit.~Comet~Tsuchinshan-ATLAS
has satisfied this condition extensively.  The phase angle reached
a maximum of 7$^\circ\!$.6 on 13~February 2023, shortly before its
second discovery.  Next, a minimum of 2$^\circ\!$.6 occurred on
30~April, followed by a maximum of 9$^\circ\!$.9 on 27~July and
another minimum of 2$^\circ\!$.0 on 30~October 2023.

In 2024, the first phase-angle maximum, of 15$^\circ\!$.2, took
place on 17~February.  Next came the already noted minimum of
2$^\circ\!$.9 on 18~April and another maximum of 30$^\circ\!$.6
on 3~July.  As the phase angle grew, it reached 15$^\circ\!$.2
on 13~May, so that the range of phase angles in late February and
most of March was the same as in the first half of May.  This
circumstance has offered an excellent opportunity to compare, as
a function of the phase angle, $\phi$, the observed magnitudes,
$H$, normalized to unit geocentric distance, $\Delta$, by
\mbox{$H_{\!\Delta} = H \!-\! 5 \log \Delta$}.

In the following experiment I apply the observations from the
same dataset that I used in Paper~1, except for the limitations
dictated by the relevant time intervals.  The CCD observations,
of excellent quality, were made by A.\ Pearce with his 35-cm f/5
Schmidt-Cassegrain\footnote{See {\tt https://cobs.si.}} and a
clear filter (Pearce 2024, personal~\mbox{communication}), so that the
CN emission was included (with obvious phase-angle implications).

In Figure~1 I plot a total of 23~values of $H_{\!\Delta}$ from the
two most recent periods of time when the phase angle was confined
to a range of 6$^\circ$ to 14$^\circ$:\ from 9~March (when the
comet was at a heliocentric distance of 3.44~AU) to 6~April (when
it was 3.09~AU from the Sun) before{\vspace{-0.02cm}} the time of
minimum phase angle (on 18~April) and from 27~April (at 2.81~AU from
the Sun) to 9~May (2.65~AU) after the time of minimum phase angle.

\begin{figure}[t]
\vspace{0.17cm}
\hspace{-0.2cm}
\centerline{
\scalebox{0.77}{
\includegraphics{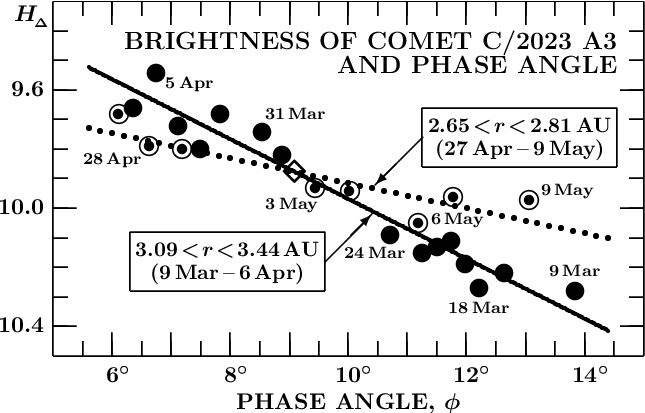}}}
\vspace{0cm}
\caption{Plot of the magnitude of comet Tsuchinshan-ATLAS normalized
to unit geocentric distance, $H_\Delta$, in the range of phase angles,
$\phi$, of 6$^\circ$ to 14$^\circ$ from the period between 9~March and
9~May 2024.  All data are from CCD observations by A.\ Pearce with his
35-cm Schmidt-Cassegrain.  The solid circles are the magnitudes before
the minimum phase angle on 18~April, the circled dots after the minimum.
The two groups of data intersect at a point marked by a diamond, at a
phase angle of 9$^\circ\!$.1 and \mbox{$H_{\!\Delta} = 9.9$}, which refers
to 29.9~March before the minimum and to 2.9~May afterwards.  Note that the
comet was 0.25~mag brighter on April~5 than more than three weeks later,
even though the phase angles on the two days were essentially the same.
Similar relations apply to the entire area to the left of the diamond,
containing the dates following 29~March in the period before the time
of minimum phase angle and the dates preceding 2~May after the time of
minimum phase angle.{\vspace{0.6cm}}}
\end{figure}

At first sight the plot looks like a compelling phase effect case,
but \ldots  The observations cover a period~of~two months, yet the
{\small \bf magnitudes {\boldmath $H_{\!\Delta}$} at any two times
when phase angles are equal do not differ from one another by
more than {\boldmath $\pm$}0.3~mag.}  For example, the comet was
practically as bright on 24~March as it was on 6~May, fully six
weeks later, even though the phase angles on the two dates differed by
less than 0$^\circ\!$.5.  And on 5~April the comet was slightly brighter
than on 28~April, more than three weeks later, even though the phase
angles were essentially identical.  So where is the comet's brightening
with decreasing heliocentric distance?

To investigate this problem more closely, I note that if the intrinsic
brightness varied as $r^{-3.5}$, the comet should have brightened
between 9~March (when its heliocentric distance, $r$, was 3.44~AU)
and 9~May (2.65~AU) by 1~mag.  Even though the phase angle was in
fact {\it greater\/} on 9~March by nearly 1$^\circ$, Pearce's
observations show that the comet was on 9~May intrinsically
brighter by merely 0.3~mag.

The data from the period of 9~March through 6~April, before the
minimum phase angle, are fitted in Figure~1 with a mean residual
of $\pm$0.06~mag by
\begin{eqnarray}
H_{\!\Delta} & \,=\, & 8.95 + 0.101 \,\phi, \\[-0.05cm]
	     &  \; & \:\!\!\!\!\!\!\pm 0.07 \,\pm \!0.007 \nonumber
\end{eqnarray}
while the data from the period of 27~April through 9~May, after the
minimum, are with the same mean residual fitted by
\begin{eqnarray}
H_{\!\Delta} & \,=\, & 9.49 + 0.042 \, \phi. \\[-0.05cm]
             & \;  & \:\!\!\!\!\!\!\pm 0.08 \,\pm \!0.009 \nonumber
\end{eqnarray}
These two lines intersect nominally at a point marked by a diamond in
Figure~1; it is given by
\begin{eqnarray}
\phi & \,=\, & 9^{\circ\!}.1, \nonumber \\[-0.24cm]
    & \;  &  \\[-0.24cm]
H_{\!\Delta} & =\, & 9.87, \nonumber
\end{eqnarray}
the phase angle referring to 29.9~March and 2.9~May~before and
after the minimum, respectively.  So at these~two times the comet
was equally bright and was at exactly the same phase angle.  The
heliocentric distances were 3.19~AU and 2.74~AU, respectively,
implying an expected, but undelivered brightening of 0.6~mag,
positively independent of the phase law.

On certain assumptions, the slopes of the lines in Figure~1 provide
information on both the phase law and the light curve.  Suppose
the normalized magnitude varies as a phase angle,
\begin{equation}
H_{\!\Delta} \sim \phi,
\end{equation}
so that the standard light-curve equation is written in expanded
form thus
\begin{equation}
H_{\!\Delta} = H_0 + 2.5 \, n \log r + b \, \phi,
\end{equation}
where $H_0$ is the absolute magnitude (at unit heliocentric and
geocentric distances), $b$ is a constant phase coefficient, and
$n$ is the power of heliocentric distance, $r$, that describes the
rate of variation of the comet's intrinsic brightness, $r^{-n}$.
Differentiating Equation~(5) with respect to $\phi$, I get, on
the assumption that $n$ is a constant,
\begin{equation}
\frac{dH_{\!\Delta}}{d\phi} = 2.5\,n\,\frac{\partial \log r}{\partial
 \phi} + b.
\end{equation}
I replace $\partial \log r/\partial \phi$ with an average slope
\mbox{$\langle d \log r/d \phi \rangle$} of a straight line fitted
through the function \mbox{$\log r = f(\phi)$} at the times of
Pearce's observations and I find that the slope comes out to be
+0.005911~deg$^{-1}$ from the dataset before the minimum phase
angle and $-$0.003735~deg$^{-1}$ from the dataset after the minimum.
Inserting these values and those of $d H_{\!\Delta}/d\phi$ from
Equations~(1) and (2) into Equation~(6), I get respectively for
the two periods:
\begin{eqnarray}
0.101 &\,=\,& b+0.01478 \,n \;\;\;{\rm (for\:9\:Mar\!-\!6\:Apr),}\nonumber\\[-0.24cm]
      &  \; & \\[-0.24cm]
0.042 &\,=\,& b-0.00834 \,n \;\;\; {\rm (for\:27\,Apr\!-\!9\:May).} \nonumber
\end{eqnarray}
The solution is 
\begin{eqnarray}
b & \,=\, & 0.063 \; {\rm mag\:deg}^{-1}\!, \nonumber \\[-0.24cm]
  &  \;   & \\[-0.24cm]
n & \,=\, & 2.55, \nonumber
\end{eqnarray}
which shows that fitting the data requires~a~huge~phase effect, with
the coefficient $b$ about~twice~the~typical~value for asteroids
(0.030--0.035 mag~deg$^{-1}$)~and~four~to~eight
times~the~averages~implied~by~comet~dust-grain~models{\vspace{-0.03cm}}
(a standard model{\vspace{-0.04cm}} by Marcus gives
\mbox{$\langle b \rangle = 0.008$ mag deg$^{-1}$} at a
phase angle of 6$^\circ$ and 0.016~mag deg$^{-1}$ at 14$^\circ$).
Another surprise is that the light curve is much more flat than are the
current estimates for $n$, near 3.5.

\begin{figure}[b]
\vspace{0.65cm}
\hspace{-0.2cm}
\centerline{
\scalebox{0.735}{
\includegraphics{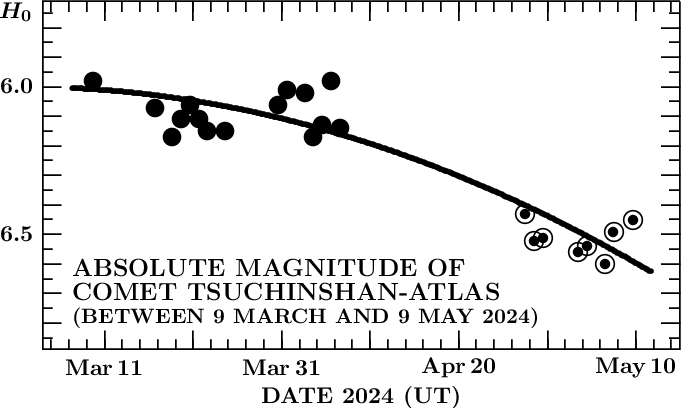}}}
\vspace{-0.2cm}
\caption{Accelerating decline of the absolute brightness of comet
Tsuchinshan-ATLAS between 9~March and 9~May 2024.{\vspace{-0.3cm}}}
\end{figure}

The final step is to insert the parameters $b$ and $n$ back into
Equation~(5) and compute the absolute magnitude $H_0$ for each
observation.  Plotted in Figure~2 as a function of time, the {\small
\bf absolute brightness was decreasing with time at an accelerating
rate,} from magnitude 6.0 near 9~March to 6.6 two months later.  An
estimated rate of fading in early May was about 0.5~mag per month.
This rate would have been higher if a less extreme value of
the phase coefficient was used.

Because the rate of brightening is known to have a tendency to diminish
with decreasing heliocentric distance, one can go a step further and
replace $n$ in the second equation of (7) with \mbox{$n \!-\! \Delta n$}
($\Delta n \!>\! 0$) and compute solutions as a function of $\Delta n$.
For{\vspace{0cm}} example, for \mbox{$\Delta n = 0.5$}~one would
get \mbox{$b = 0.061$ mag deg$^{-1}$} and \mbox{$n = 2.73$} before the
phase angle minimum and 2.23 after it.  A decreasing steepness of the
light curve appears to have a rather insignificant effect on the
phase coefficient.

\section{Discussion and Conclusions}
In an effort to understand the role of the phase law on the absolute
magnitude of comet Tsuchinshan-ATLAS, I examined its light curve when
the phase angle varied between 6$^\circ$ and 14$^\circ$.  The first
period extended from 9~March to 6~April 2024 (3.44--3.09~AU from the
Sun), before the minimum phase angle was reached on 18~April, the
second period from 27~April to 9~May (2.81--2.65~AU) after the
minimum.

The choice of the lower boundary of 6$^\circ$ was determined primarily by
the need of having a sufficient gap between the two periods (I adopted
three weeks), while the choice of the upper boundary was dictated by the
maximum near 15$^\circ$ in mid-February.  The phase angle span of 8$^\circ$
was enough to closely inspect the near-pairing of a number of the data
points.  At an equal phase angle, the comet's brightness before and after
mid-April, was found, after applying correction for the geocentric
distance, to be nearly equal, implying intrinsic fading.

Comparison of the rates of brightness variations with the phase angle
in the periods before and after the time of its minimum allowed me to
determine the phase coefficient and the rate of variation with
heliocentric distance.  The solution offered rather unusual results:\ a
very steep phase law, with the coefficient exceeding 0.06~mag~deg$^{-1}$,
and a light curve more flat than expected, with \mbox{$n = 2.55$}.\,\,

Inserting the parameters into the equation of the light curve led
eventually to the determination of the absolute magnitude as a
function of time over the entire investigated period.  The result
was a clear systematic decrease in the absolute brightness from
magnitude 6.0 in early March to 6.6 in early May, with an accelerating
rate, equaling 0.5~mag around 9~May.  If this trend has continued, the
absolute magnitude would be close to 8, possibly fainter, at the time
of this writing (mid-July).  With \mbox{$\Delta = 2.03$ AU}, \mbox{$r
= 1.67$ AU}, and \mbox{$\phi = 30^\circ$} on 14~July, the comet would
be close to apparent magnitude 13.  Instead, it is observed near
magnitude 10. 

Indeed, the comet was recently reported to brighten (Pearce, personal
\mbox{communication}) just as the phase~angle was peaking at 30$^\circ\!$.6!~The bizarre~law~and~this~remarkable anticorrelation suggest that the
phase~\mbox{phenomena} have little effect.  Rather,
fluctuations~linked~to~\mbox{incidents} of {\it increased episodic activity\/}
dominate~the~comet's~perfor\-mance.  The comet may be experiencing another
surge, as a new cluster of fragments got activated~for~a~short~period of
time following their separation from~the~nucleus.\\[-0.2cm]
%
%
\begin{center}
{\footnotesize REFERENCES}
\end{center}
\vspace{-0.45cm}
\begin{description}
{\footnotesize
%
%
%
%
%
%
\item[\hspace{-0.3cm}]
Marcus, J.\ N.\ 2007, Int.\ Comet Quart., 29, 39
\\[-0.35cm]
\item[\hspace{-0.3cm}]
Sekanina, Z. 2024, eprint arXiv:2407.06166 (Paper 1)}
%
\vspace{-0.55cm}
\end{description}
\end{document}